\documentclass[aip,jap,amsmath,amssymb,showpacs,twocolumn]{revtex4}
\usepackage{graphicx} 
\usepackage{dcolumn} 
\usepackage{array} 
\usepackage{bm} 
\usepackage{latexsym} 
\usepackage{epstopdf}

\begin{document} 
\title{Coexistence of  bunching and meandering  instability in simulated growth of 4H-SiC(0001) surface}
 \author{Filip Krzy{\.z}ewski} 
\email{fkrzy@ifpan.edu.pl} 
 \address{Institute of Physics, Polish Academy of Sciences,
Al. Lotnik{\'o}w 32/46, 02-668 Warsaw, Poland} 
\author{Magdalena A. Za{\l}uska--Kotur }	
\email{zalum@ifpan.edu.pl}
 \address{Institute of Physics, Polish Academy of Sciences,
Al. Lotnik{\'o}w 32/46, 02-668 Warsaw, Poland and Faculty of Mathematics and Natural Sciences,
Card. Stefan Wyszynski University, ul Dewajtis 5, 01-815 Warsaw, Poland}
\begin{abstract}
Bunching and  meandering  instability of steps at the 4H-SiC(0001) surface is studied  by the  kinetic Monte Carlo  simulation method. Change in the character of step instability  is analyzed  for  different   rates of  particle jumps   towards  step. In the experiment   effective  value of jump rates can be  controlled by impurities or other growth conditions. 
An anisotropy of jump barriers at the step  influences 
the character  of surface structure formed in the process of  crystal growth. Depending on the growth parameters different surface patterns are found. We  show phase  diagrams of  surface patterns  as a function of temperature and crystal growth  rate for two different choices of step kinetics anisotropy. 
Jump rates  which effectively model high inverse Schwoebel barrier (ISB) at steps lead either to regular, four-multistep or bunched structure. For weak anisotropy  at higher temperatures or for lower crystal growth rates meanders and  mounds are formed, but on coming towards lower temperatures and higher rates we  observe bunch and meander coexistence. 
These results show  that interplay between simple dynamical  mechanisms induced  by the asymmetry of the step kinetics  and  step movement assisted by the step edge  diffusion are responsible for different types of  surface morphology.
\end{abstract}
\date{\today}
\maketitle
\section{Introduction}
Silicon carbide (SiC) is intensively studied material due to its application in  high temperature, high power and  high frequency  electronic  devices.    Lately it became even more interesting as a basis for graphene production. Being the matter of  continuous interest it is the subject of many experimental as well  theoretical investigations.  Depending on growth conditions many various step patterns as multisteps,  bunches or meanders at  SiC surface were seen. 
The morphological instability of step trains during growth of  6H-SiC(0001) \cite{ohtani,ohtani2,matsunami,matsunami2,matsunami3,omar,syvajarvi,yamamoto} and 4H-SiC(0001) \cite{Kimoto,yazdanfar,dong,K,C,LV} surfaces were studied in various experimental conditions. It was observed that such instability can be impurity induced and the mechanism of this influence can be related to the change of the effective jump barriers at the steps \cite{ohtani,ohtani2}.

Complex morphological features at the growing crystal surface follow from an interplay among nucleation, diffusion,
and the incorporation of adatoms at steps \cite{sato,misbah,yagi}. It was shown that the balance between the flux of adparticles attaching step from the upper terrace and flux from the lower terrace has key impact  on the step stability and 
the emergence of the  final surface pattern\cite{misbah,schwoebel,bales}. 
During crystal growth or sublimation process various factors may affect the fluxes. The most often discussed  reason for imbalanced fluxes   is so called Schwoebel barrier (SB) at steps \cite{schwoebel}. Similar effect onto the step stability has the  step movement during crystal growth or sublimation process \cite{Dufay,stoyanov1}. The inverse Schwoebel barrier (ISB) acts in the opposite direction that usual SB slowing  flux that comes towards the step from the lower terrace and it can stop or invert  the  effects of the  natural asymmetry in step dynamics. These two factors: ISB and step movement  during crystal growth induce particle fluxes in the opposite directions.  Step movement  appears to be  quite effective mechanism to generate step bunching in the process of crystal annealing \cite{Dufay,stoyanov1,zaluska1} and  the presence  of   ISB  is an origin of  bunches during  crystal growth \cite{sato}. SB is  known as a reason for step meandering instability at  surfaces of grown crystals \cite{misbah,bales},  however the step  meandering due to  the diffusion along step edges either via kink Schwoebel barrier \cite{murty,pierre,politi} or via unhindered step-edge diffusion \cite{nita}  often overcomes SB effect.

 All particle fluxes either along surface  or along  step edges depend on  the growth parameters like temperature and growth rate. They follow in different way changes of parameters and as an effect various surface patterns are created. The analysis of experimentally observed bunching process at  SiC(0001) surface indicates that the impurity  adsorption influences  the final step pattern of growing crystal \cite{ohtani,ohtani2}.  It was argued that the  nitrogen adsorbed at steps can enhance the incorporation rate of adatoms from the upper
terraces or reduce that from the lower terraces \cite{ohtani,ohtani2}, thus changing the value of effective ISB.
 We show concrete examples  of   step patterns that result as a competition of ISB and step rate  at different growth conditions. Temperature - rate pattern diagram changes for   different choices of  ISB height.

 Evolution of  4H-Si(0001) surface is modeled by use of kinetic Monte Carlo simulations  \cite{borovikov,camarda,camarda2,camarda3}
and    bunching - meandering  instability is studied as a function of temperature and the  crystal growth  rate  for different ISB values. 
We show that with  high ISB value steps create bunches at low temperatures which depending on the step rate changes into 4-step patterns or regular step arrangement. When ISB is  absent  no step bunching happens and only meandered, 4-step and mound structures are build.
The situation is different for  low ISB values. Depending on the  temperature we observe  4-step structures or regular in-phase meanders, which change into mounds or    bunch  and meanders coexistence for higher step rates. This last  structure  is very similar to the one seen experimentally in Ref. \onlinecite{neel} and obtained in Ref. \onlinecite{yu2,yu} by phase-field approach. It is  simultaneous bunch and meandered structure, an intermediate ordering between bunches and meanders. In our case we realize such structure   by the presence of ISB, which is not so  large  to win over the  flux up the steps in the whole area, but large enough  to cause step bunching in some parts of the system. 

Our  kMC  model is described in Sec I, then we discuss results of  simulations in Sec II  for high  ISB and low ISB successively  and summarize by conclusions in Sec III.

 \section{The model }
Kinetic Monte Carlo (kMC) simulations were carried out on the  lattice  of 4H politype of silicon carbide. Elementary cell of such a crystal consists of eight  alternating  layers  of  Si and  C atoms.  Consecutive double SiC layers of the lattice are shifted towards each other in order to form ABAC stack which corresponds to the silicon face of polar 4H-SiC crystal structure \cite{borovikov,camarda,camarda2,camarda3}. 
The surface of  4H SiC(0001) symmetry was modeled. Both silicon and carbon atoms are controlled  in the simulation.  The energy of grown crystal  includes interactions between both types of adatoms
\begin{equation}
H=- E_{\rm SiC}\sum_{\rm NN} n_i n_j - E_{\rm SiSi} \sum_{\rm NNN} n_i n_j -E_{\rm CC} \sum_{\rm NNN} n_i n_j
\end{equation}
where $n_i=0,1$ means  site  occupation and sums are over nearest neighboring atom pairs (NN) and next nearest neighbors  (NNN). NN bonds are  between silicon and carbon atoms and  NNN bonds correspond to C-C and Si-Si ones. Energy constants in the formula above are  some effective values and  can  be determined by the analysis of system behavior and by comparing with experimental data for concrete system. We used the same  constants  that  were assumed  in Ref.\onlinecite{borovikov}, where some characteristic features of SiC(0001)  kinetics were reproduced.   Thus we have  $E_{\rm Si C}=0.75 eV$ and $E_{\rm CC}=0.65 eV$ and $E_{\rm Si Si}=0.35eV$ respectively. Every atom has up to four NNs, which  lay in neighboring layers and it is bound  with up to  twelve NNN. Six of them lay in the same layer, next three one  layer  below and the last three in the layer above.  Geometry and directions of bonds are given by positions of atoms at A, B or C layers forming 4H-SiC crystal.

At the beginning of every simulation system consists of $N_s$ steps of equal width $W_s$ which sets the initial miscut. Steps ended by silicon alternate with the ones with carbon on the top. The surface is misoriented in $[01\bar{1}0]$ direction. In the system we set periodic boundary conditions at edges perpendicular to steps and helical boundary conditions for edges parallel to the steps.  In order to close such a boundary condition in the correct way, step number $N_s$ must be divisible by 8. That is the number of monoatomic layers at 4H-SiC elementary cell and also the minimum number of steps to be considered during simulations.

The first step  of simulation  is an  adsorption of silicon and carbon atoms. Probability of such a process is equal to the external particle flux $F$, $p_A=F$.
In the next step jump directions  for each particle at the surface are chosen. Then the probability of a particle diffusion  $p_J$ is calculated. It depends on the system temperature $T$ as well as on initial $E_i$ and final $E_f$ energy of jumping atom  calculated as follows
\begin{equation}
\label{E}
E_{i,f}=E_{\rm SiC}\sum_{(NN)_{i,f}} n_{NN}+E_{ \rm XX}\sum_{(NNN)_{i,f}} n_{NNN}.
\end{equation}
In above formula $(NN)_{i,f}$ and $(NNN)_{i,f}$ correspond to nearest  and next nearest  neighbors of the initial or final particle positions  respectively and XX=C C when energy of C atom is calculated and XX=Si Si  for Si particle.When the initial energy is higher than the final, the atom jumps from deeper potential well to the shallower one. In this case
\begin{equation}
\label{p_J}
p_J=\nu \exp(-\beta(E_i-E_f+\Delta_X)),
\end{equation}
where $\nu$  is attempt frequency, and $\beta=1/k_BT$, $k_B$ is the Boltzmann constant and $\Delta_X$ with  corresponds to the diffusion barrier for C or Si atom ($X$=C or Si).  
 In the case when the initial energy of interactions is lower than the final, hence the atom jumps from shallower to deeper potential well $p_J=\nu \exp(-\beta \Delta_X)$. 
  According to several studies   barrier for diffusion of  carbon  atoms is higher \cite{diff1,diff2,borysiuk}.   We 
assumed the  difference in the diffusion barrier as 0.3 eV.   Thus  we have $\Delta_C=0.3eV$ for carbon and $\Delta_{Si} = 0$ for silicon adatom. Such assumption  result   in the carbon mobility few times slower than this for silicon adatoms.
Silicon diffusion over the surface  is the fastest process and sets the time scale of the whole studied kinetics. All other processes are set up relatively to this fastest  jump and on assuming that $\nu=10^{11} ML/s$ (monolayer per second) and the barrier for  Si diffusion is around $0.8 eV$   we can  determine extend of the studied below  growth rate, given by $F$.
We studied fluxes  $F$ within  the range of 10 ML/s up to $10^4$ ML/s (around 50-5 $^. 10^4 \mu$m/h).
In such a way we quickly enter area  of  growth rates which are too high for the experiment. However it is worth noting that all numbers used here are only rough approximation of the real values, and  they have their meaning only in  qualitative description of  behavior of grown   crystal. Each of approximated  value of parameter can be in fact  moved toward higher or lower values, according to some additional data,  thus changing range of studied rates and temperatures. 

In order to model ISB value an additional barrier $B_I$ is added for jumps towards and out of  neighboring sites.
In such a case diffusion close to step edges is slowed down and  a  formula for the  jump probability changes into 
\begin{eqnarray}
\label{p_J_BI}p_{ISB}&=&\nu  \exp(-\beta( E_i-E_f+\Delta_X+B_I)  \nonumber \\
p_{ISB}& =& \nu  \exp(-\beta( \Delta_X+B_I) \ \ \ \ \ \ {\rm for} \ \ \ \ \  E_i >E_f
\end{eqnarray}
In such way  incorporation rate of adatoms from upper terrace is different than that from the lower one \cite{uwaha}. Setting an  additional barrier at the lower terrace is  more  effective method to control particle to step fluxes than reduction of barrier at the higher terrace, as it  was discussed  in Ref \onlinecite{xie}  and what we have also checked in our simulations.

Each particle at the terrace can desorb from the surface. Probability of that process depends on initial energy of desorbing atom and desorption barrier $\mu_X$ of silicon $X=Si$ or carbon $X=C$:
\begin{equation}
\label{p_D}
p_D=\nu_{des} \exp(-\beta(E_i+\mu_X)).
\end{equation}
In the simulations below we assume $\nu_{des}=\nu$ and very low desorption rate $\mu_X=10 eV$. Particles practically do not leave the surface.

Simulations begin from the configuration of parallel, equally distanced steps of carbon and silicon layers. Below results for the surface with miscut of $8^o$ are presented. For each studied system we run simulation until several hundreds of layer grow on top of the crystal and observe the shape of stationary pattern that results from the surface kinetics. Different types of the surface patterns found  on studying the system  kinetics are organized in  the temperature - growth rate phase diagram.

\section{Bunching and meandering step instabilities. }
\subsection{Mechanisms of  step instabilities}
As long as the particle flux from the upper terrace is balanced with the flux from the lower terrace
regular pattern of straight steps is a stable configuration\cite{misbah,bales,uwaha}.
This situation changes however  when these two fluxes achieve  different values. 
Step movement  forward during crystal growth or  backward during crystal annealing  can be the first, natural  reason for imbalanced fluxes. Step movement during crystal growth induces  advection flux up the steps and similarly   step movement during crystal sublimation gives advection flux down the steps. 
 Crystal growth or sublimation are slow processes and so step  rate is slow  as compared with the surface diffusion and often it is neglected in the calculations.   However as it was shown in Ref. \onlinecite{Dufay,stoyanov1} 
  the asymmetry caused by it is comparable to the one  in the    electromigration process and it  is  also enough to generate step bunching.  On the other hand step movement forward  during crystal growth   can be natural source of the meandering step instability. Meandering instability  however differently than bunches happens in two dimensions and diffusion along step edges rather than  surface diffusion can be here the main source  of the  kinetic asymmetry \cite{murty,pierre,politi,nita}. It acts via Schwoebel barrier  at kinks or via unhindered step-edge diffusion.  In our simulations  of 4H-SiC(0001) we observe  meandering step structure even for  low rates of  crystal growth, so it seems that step edge diffusion is responsible for  creation of such structures. Moreover, we see that steps stay straight for higher temperatures at the same growth rates what suggests the presence of kink Schowebel barrier. The presence of kink Shwoebel barrier can be attributed to the bonds that have to be broken when particle overcomes step kink. For higher temperatures  barrier of given height results in lover differences of particle fluxes. 

The particle flux balance changes when some additional factors leading to the kinetic asymmetry  are present in the system.
\begin{figure}
\includegraphics[width=8cm]{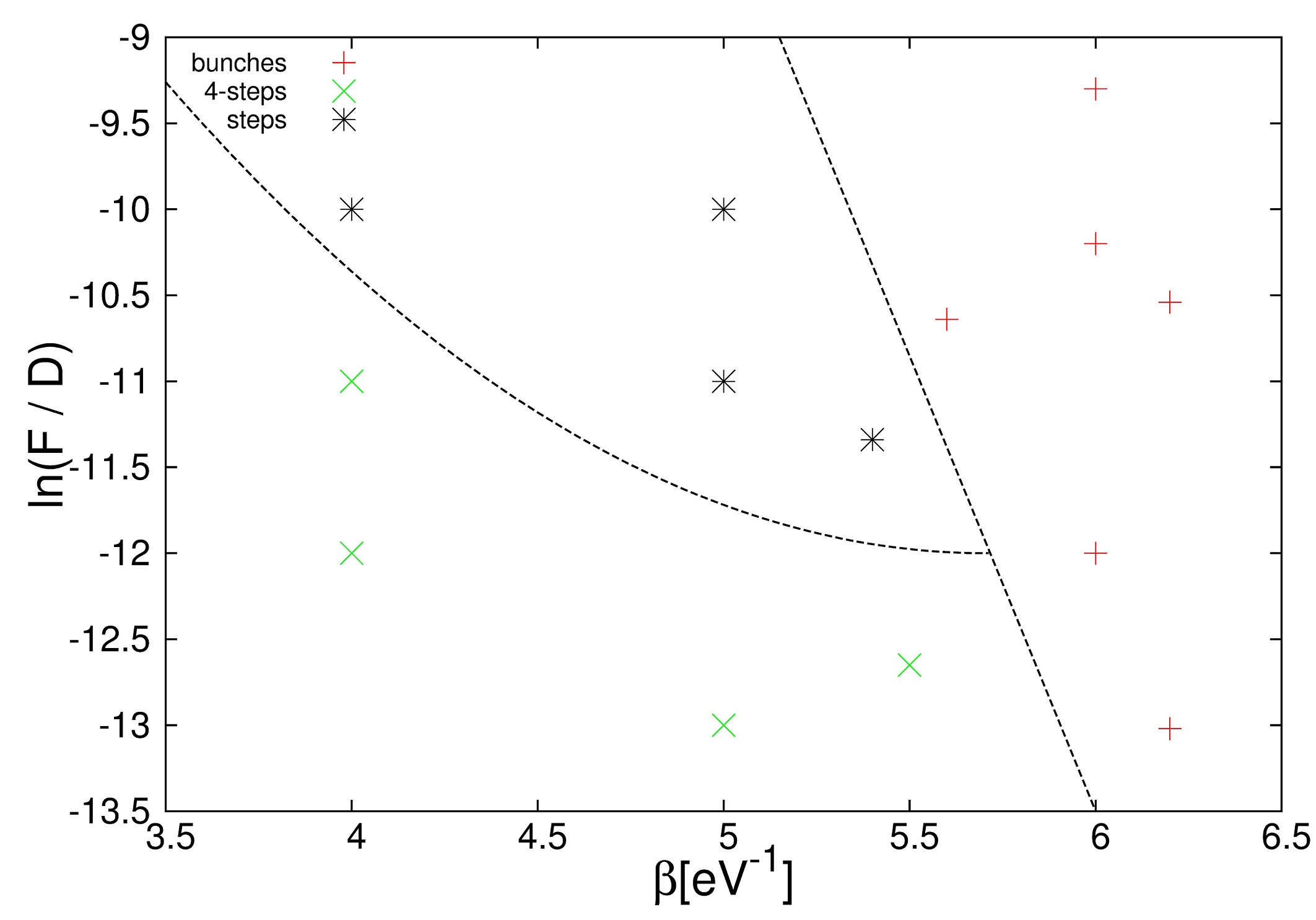}
\caption{\label{ph2} Diagram of different step patterns  at SiC(0001) surface  in the $F/D$ and $\beta$ plane calculated for  the case when  ${E_B}^{Si}={E_B}^C=0.75eV$}
\end{figure}
\begin{figure}
\includegraphics[width=7cm]{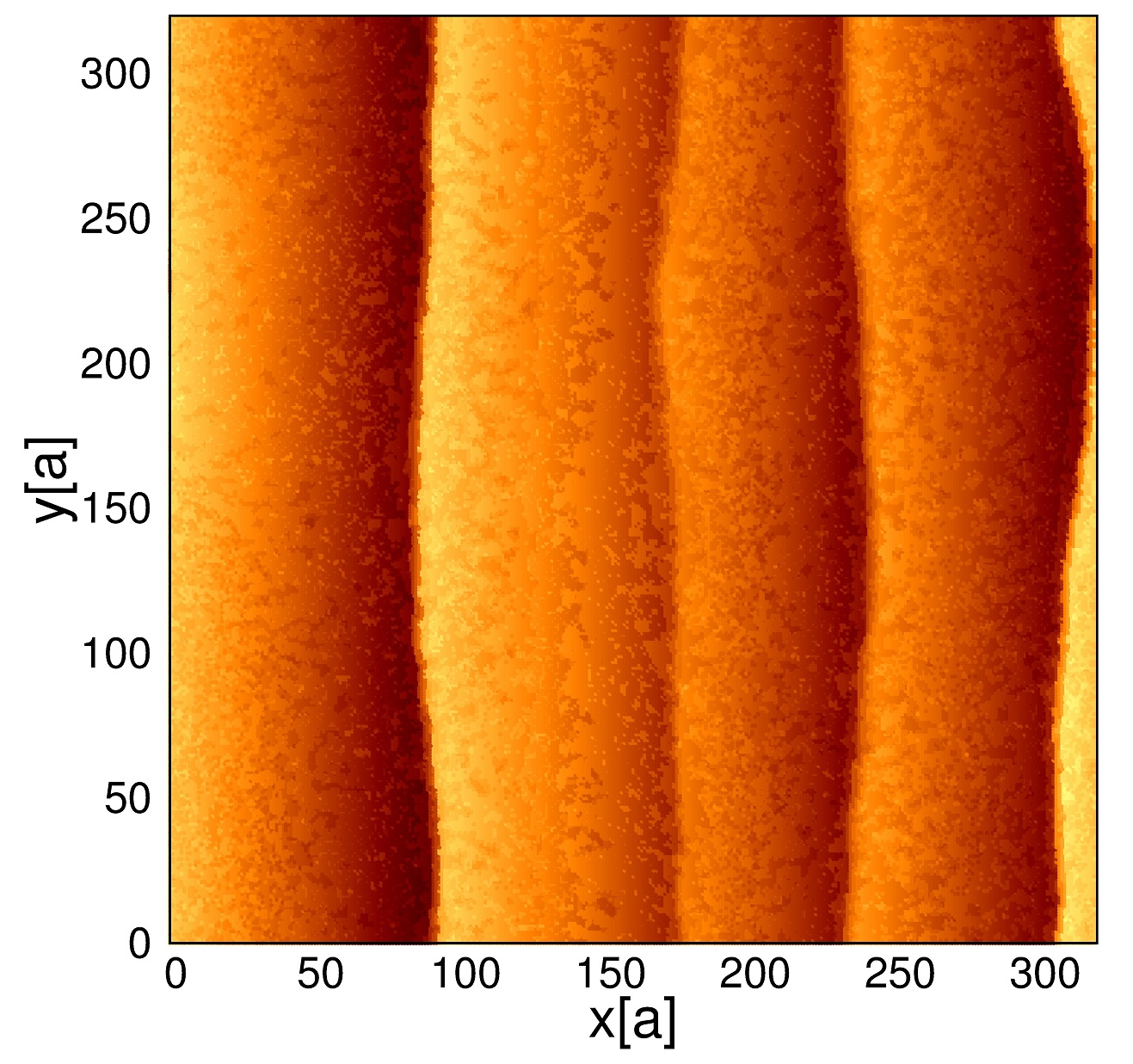}
\caption{\label{bunch} Bunches at SiC surface for $\ln(F/D)=-10$ and $\beta=6 eV^{-1}$}
\end{figure}
The most often studied  source of  imbalanced fluxes at the step  is the SB - difference in barriers for the jump to the step from the upper and lower terrace. It means  some additional barrier for diffusion at the upper step  side \cite{uwaha,schwoebel,xie}. For growing crystal  such barrier causes step meandering, so it enhances all effects observed for systems without any step barriers. In order to invert this tendency and induce step bunches instead of step meanders in the growing system ISB is  needed \cite{sato,xie}.  Such barrier   increases particle flux from the upper step side on comparing  with the flux from  the lower terrace. 
The anisotropy like that can be modeled in Monte Carlo simulation procedure either by decrease  of the barrier for the diffusion of particles from the upper terrace or  by increase of  the barrier at the lower terrace. The second method is more effective one as it was discussed in Ref. \onlinecite{xie}. We have checked this for our systems in practice on comparing  data for both realizations of particle dynamics process.  The result of our simulations was that only the second method leads to the bunched patterns, whereas the first one is not. 
\begin{figure}
\includegraphics[width=7cm]{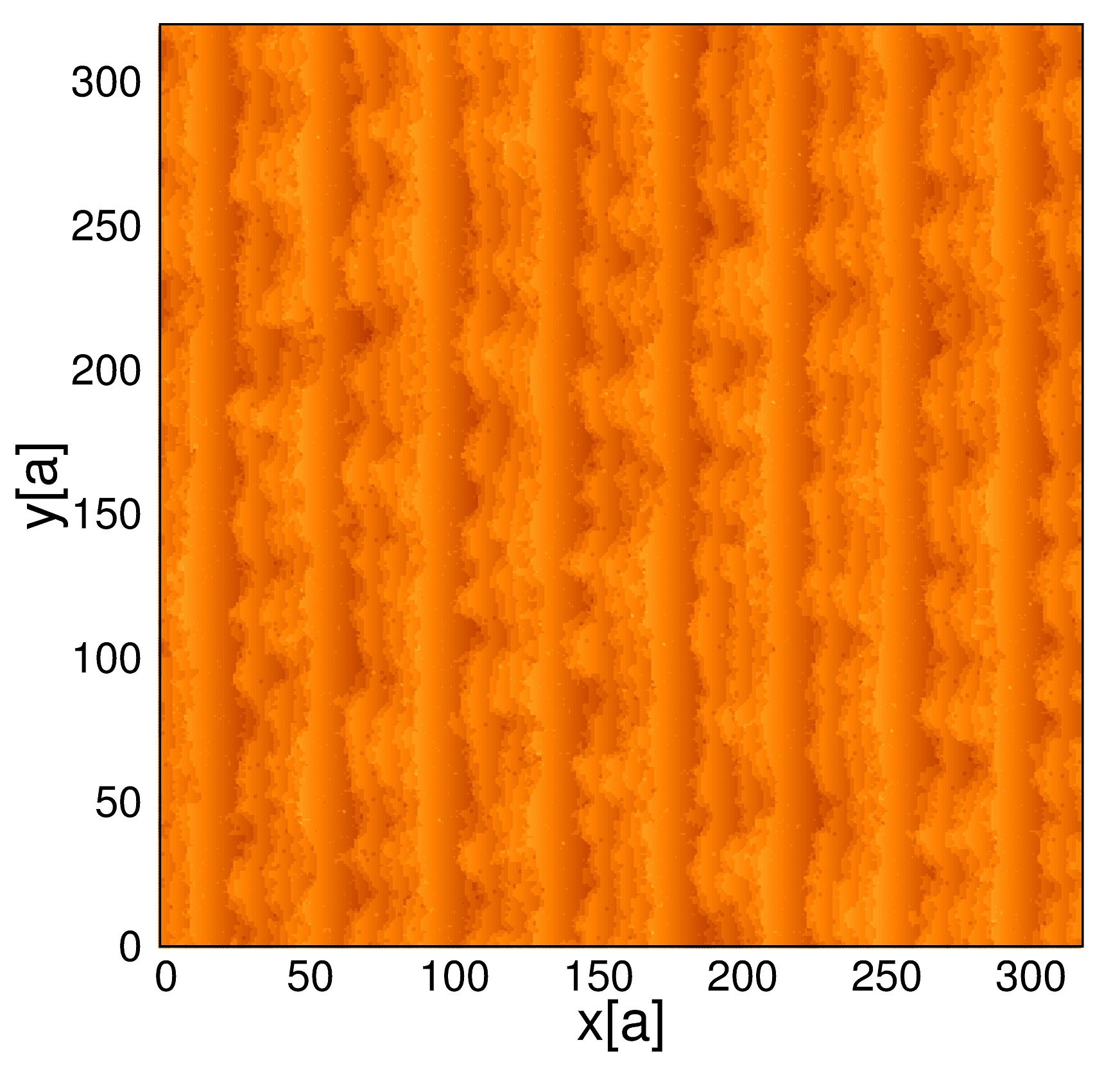}
\caption{\label{four} 4-step structure at SiC surface for $\ln(F/D)=-12$ and $\beta=4 eV^{-1}$}
\end{figure}
Hence in our model we  use formula (\ref{p_J_BI}) for the particles that  attach step from the lower terrace and in such a way we induce flux, which moves in opposite direction to the  step-flow advection of  particles  during crystal growth process.

Particle fluxes along step edges and along surface can be larger or smaller depending  on the step velocity, which is related to  the external particle flux F and to the surface miscut. On the other hand intensity of the opposite particle flux due to the ISB asymmetry  depends on the temperature. For higher temperatures, factor $\beta$ is lower and the ratio of jumps at both sides of the step decreases. Due to competition  of both  asymmetry factors we can expect different surface behavior  for different temperature and particle  fluxes. As it was discussed in Refs \onlinecite{ohtani,ohtani2} the presence of  impurities of given type can  induce or remove additional diffusion barrier, hence we expect that  various values of effective jump barriers at the step can be realized at different growth conditions. We have checked  the behavior of the model for two different diffusion barriers $B_I$ and indeed several regions of different step patterns can be found in  the $F/\nu$ and $\beta$ plane.

\subsection{Step patterns in the system with high ISB} 
Let us first discuss the case where  high ISB  was assumed   for both Si and C adatoms $B_I=0.75eV$.  Large ISB values have been  found in experimental systems, for example 1eV at Si surface \cite{Si}.
Resulting diagram of different surface structures at 4H Si(100) is presented in Fig \ref{ph2}. 
On changing flux and  temperature three different step arrangements can be observed . At low temperatures (high $\beta $ values) ISB is the dominating factor and leads to the step bunching process. Steps group in bunches which join together for longer times of crystal growth \cite{misbah,omi2,stoyanov,sato1}.
The example of such step pattern is shown in Fig. \ref{bunch}. Bunches are rather straight. It can be seen that they can bend slightly during their evolution. At low temperatures bunch structure was observed for all studied particle fluxes, whereas  for higher temperatures  two different types   of straight step structures  were found. When crystal is grown   slowly  it  leads to a regular four step  structure, what means that we have here stationary  pattern of one unit cell multi-steps .  Example of such structure is shown in Fig. \ref{four}. Difference between regular and shallow four-step structure and large bunches is clearly visible when we compare Fig \ref{bunch} and  Fig \ref{four}. For higher rates of crystal growth at higher temperatures we enter region   of regular steps at the surface.  This pattern changes its character  into the rough surface when  growth becomes faster. Within all parameters range no meandering was seen except small ripples at three out  of four  steps in 4-step structure in Fig \ref{four}. This means that high ISB, which leads to the emergence of straight bunches damps step meanderings at the same time. One of reasons that meandering instability is suppressed is the presence of ISB  at kink, the same as  ISB at steps. The presence of this barrier can stop the fluxes that flow along step. Similarly we do not see any domains build at terraces. This is again caused by ISB effect.
 \begin{figure}
\includegraphics[width=8cm]{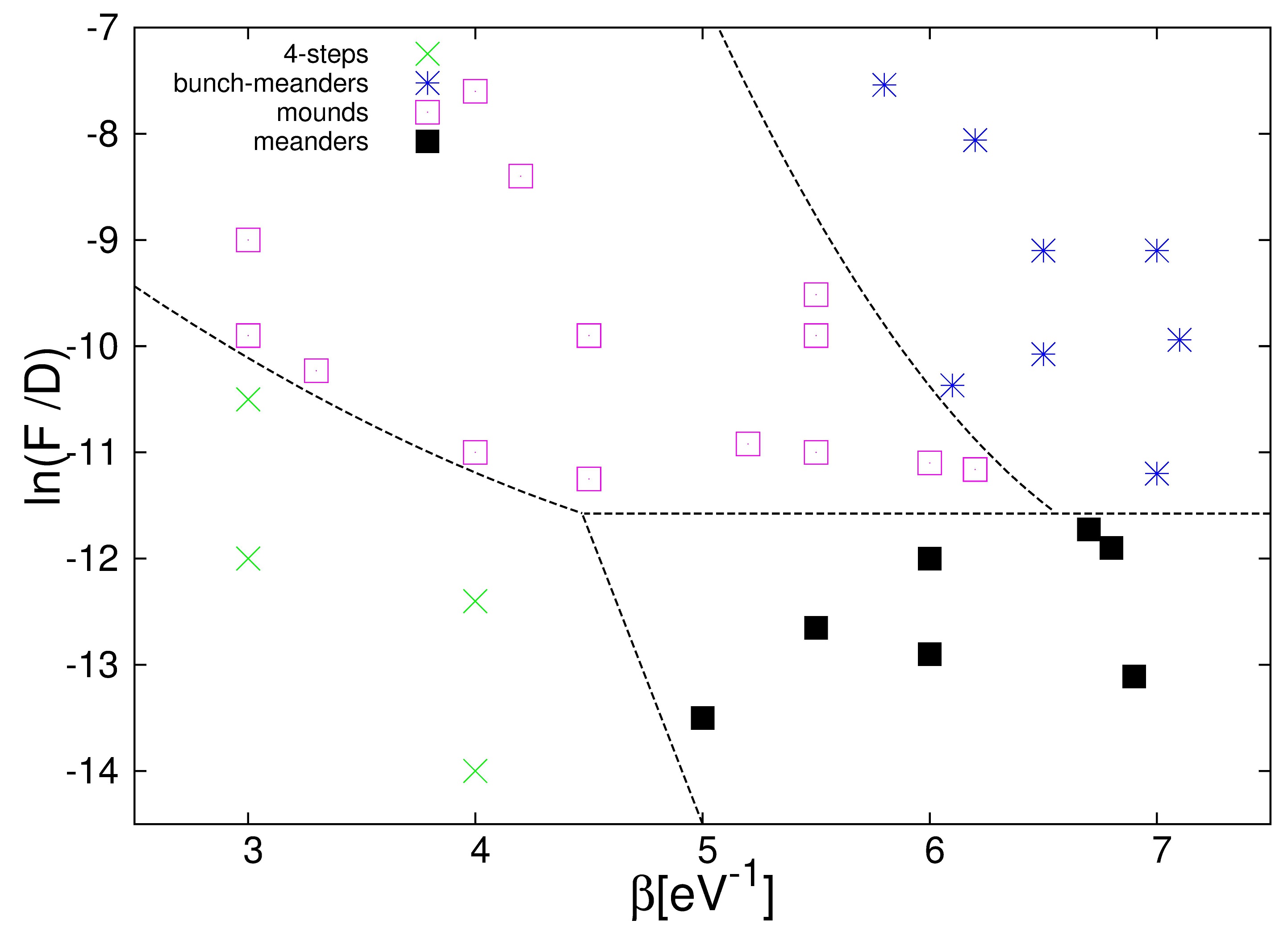}
\caption{\label{ph1} Diagram of different step patterns found at  SiC(0001) surface  in the $F/D$ and $\beta$ plane calculated for  the case when  ${E_B}^{Si}=0.5eV$ and ${E_B}^C=0.2eV$ }
\end{figure}
\begin{figure}
\includegraphics[width=7cm]{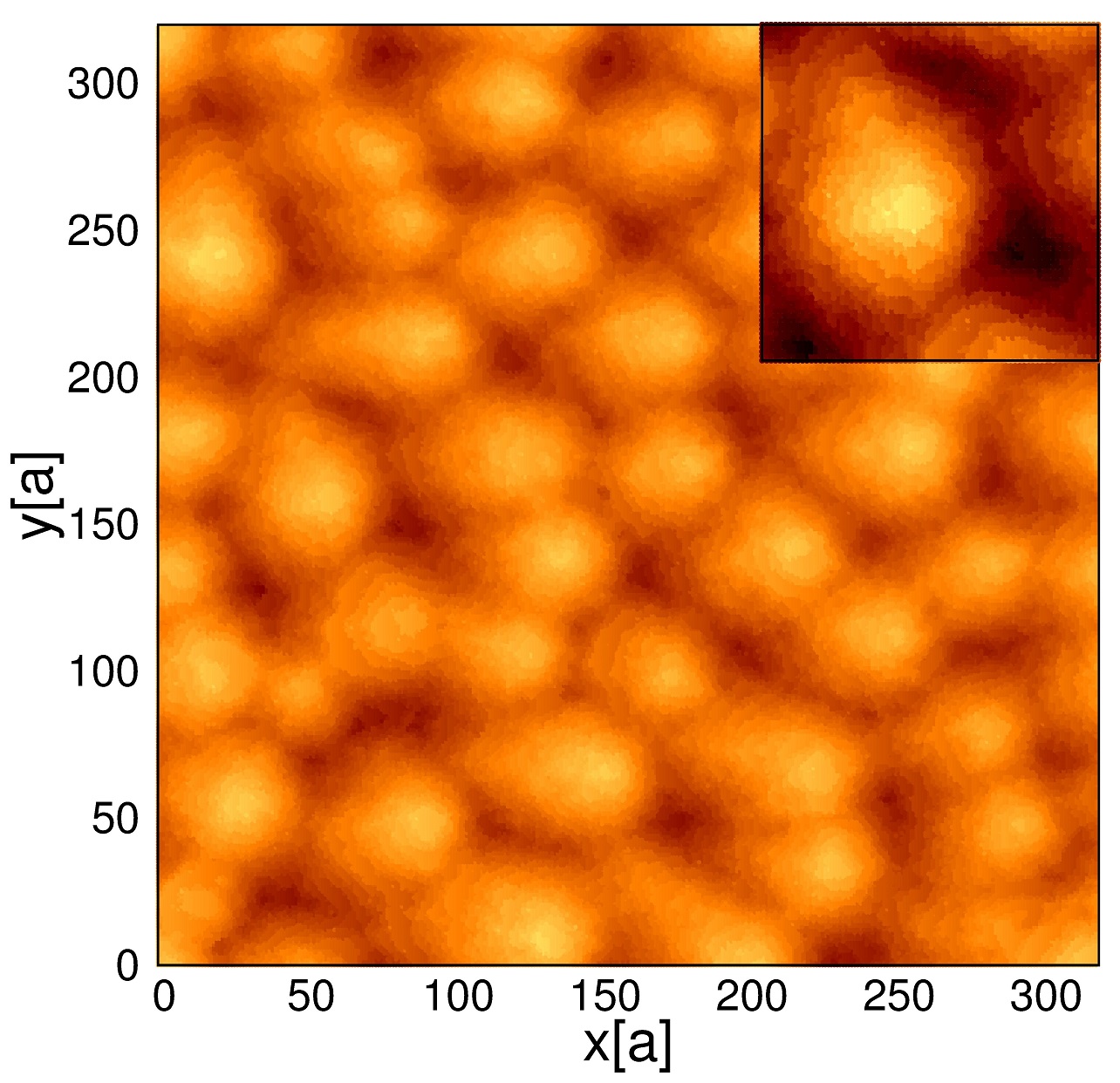}
\caption{\label{mound} Structure of mounds at  SiC surface for $\ln(F/D)=-9.9$ and $\beta=4.5$. Insert shows one of mounds  in close-up.}
\end{figure}
\begin{figure}
\includegraphics[width=7cm]{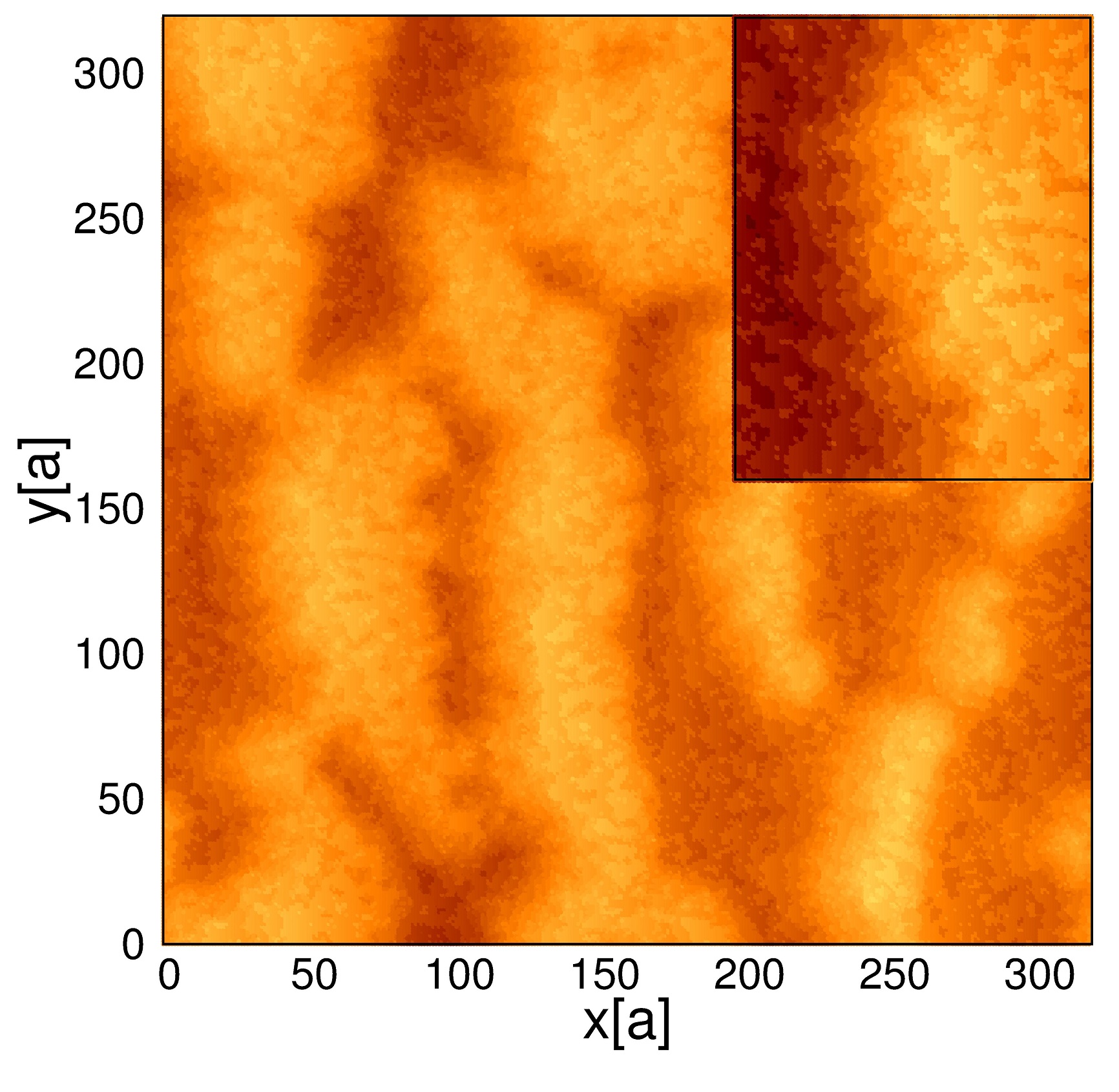}
\caption{\label{b-m} Meandered and bunched structure at SiC surface for $\ln(F/D)=-9$ and $\beta=6.5$. Insert shows part of the picture  in close-up. }
\end{figure}

\subsection{Step patterns in the system with low ISB}
\begin{figure}
\includegraphics[width=8cm]{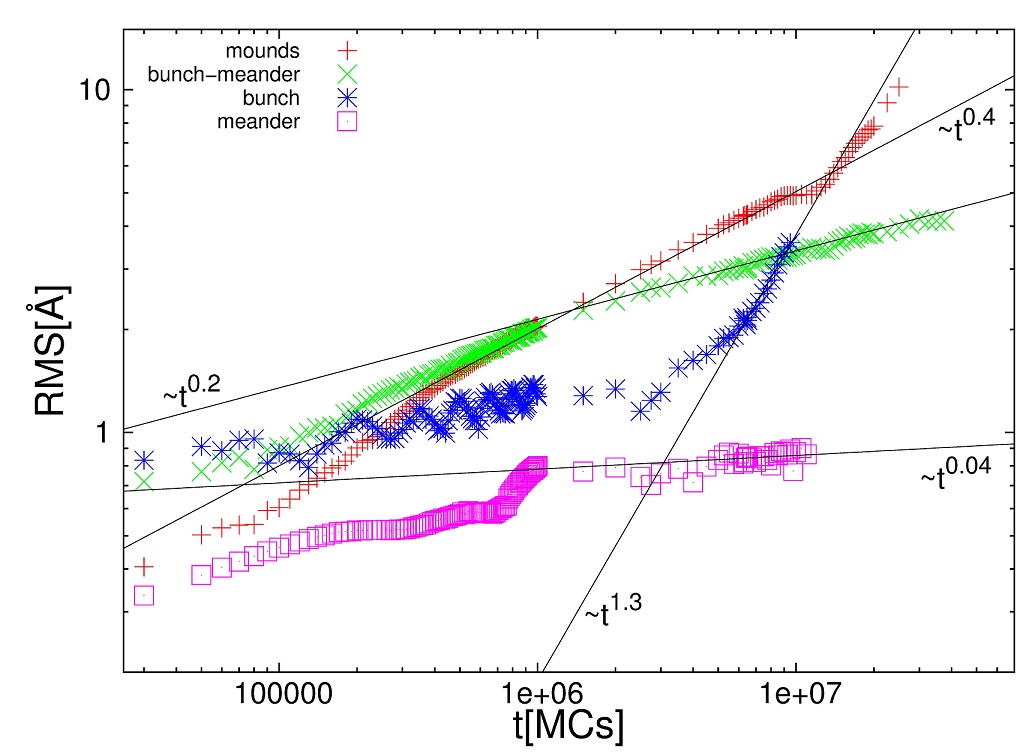}
\caption{\label{rough} Surface  roughness as a function of time for different patterns, measured as root mean square of height  of the surface. Successive curves describe surface patterns presented in Figs 5,6,2  and 11.}
\end{figure}

 The same range of growth parameters was studied for the system of lower values of ISB.  In this case we also assumed that two barriers are different and so we have $B_I=0.5eV$ for C atoms and $0.2eV$ for Si atoms. In Fig. \ref{ph1} we show   diagram  of patterns obtained  for this  system.  The character of this diagram does not change as long as both barriers are  lower than     $0.7 eV$ and stay around  value $0.4eV$.  Within the studied range of parameters four different  step structures  can be found.  For fast crystal growth at higher  temperatures mound formations can be seen.  In Fig \ref{mound} characteristic  round mound structures are clearly visible. When temperature decreases the structure of  mounds rapidly changes and we can see elongated structures of meandered bunches (Fig \ref{b-m}).  The difference between mound and this last ordering  is apparent. The first are round regular structures, the second are elongated structures,  bunched along steps. 
Mounds increase, glue together and as an effect roughness of the surface increases faster than  this of bunch and meandered structure. The difference between roughness  as a function of  time for bunch and meandered structure and for mounds  can be seen in Fig. \ref{rough}. Presented value was calculated as a root mean square  of height $h_i$ correlation function  $RMS=\sqrt{ <(h_i-h_0)^2>}$, where $h_0$ is mean surface height. It can be seen that roughness of  mounds increases faster.   Also the growth is not steady, roughness jumps up when some  reorganization covering a large area happens at the surface.  
\begin{figure}
\includegraphics[width=8cm]{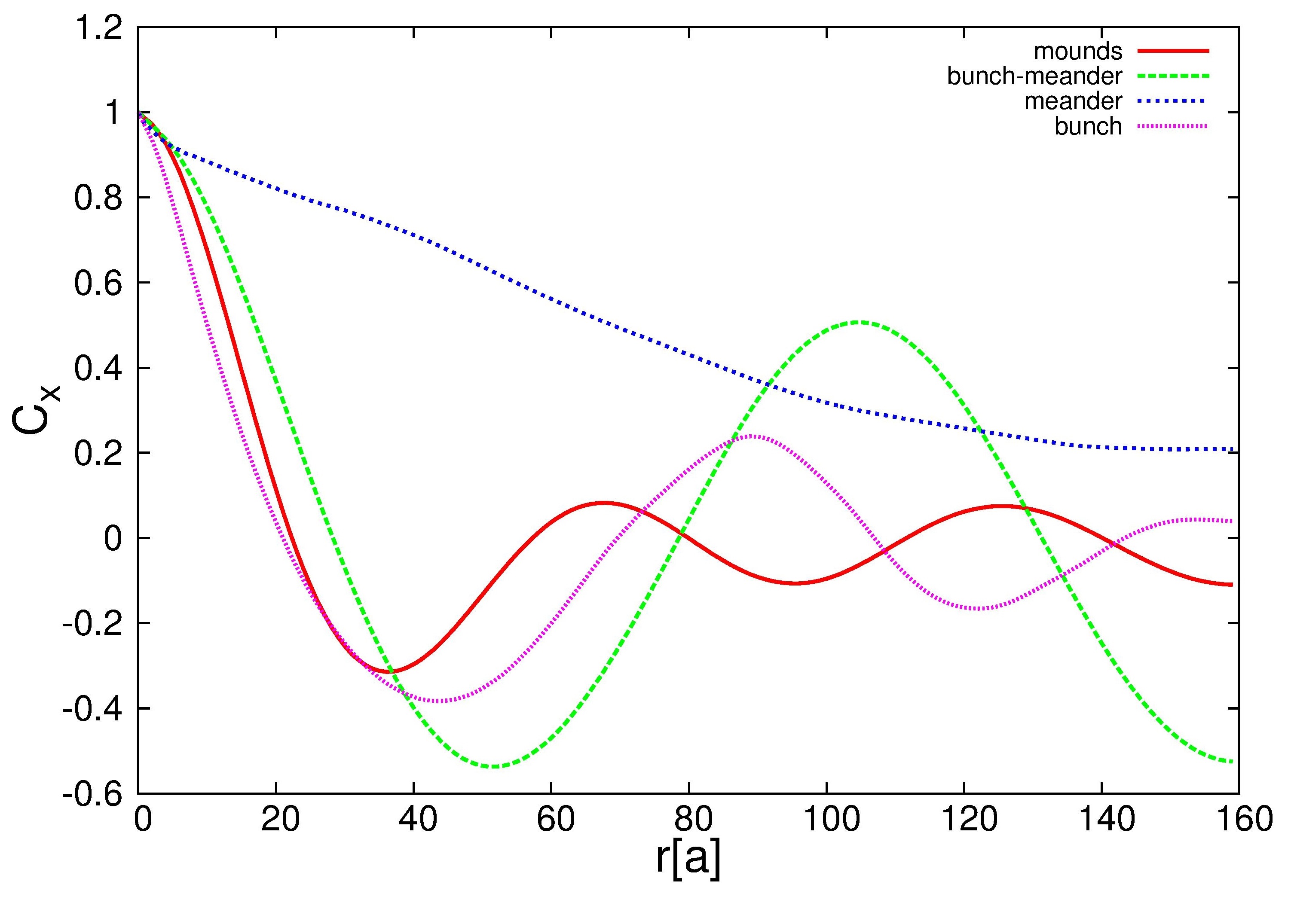}
\caption{Correlation function along $x$ direction for the same surfaces as in Fig. 7 plotted at $t=10^7$.
\label{cx}}
\end{figure}
\begin{figure}
\includegraphics[width=8cm]{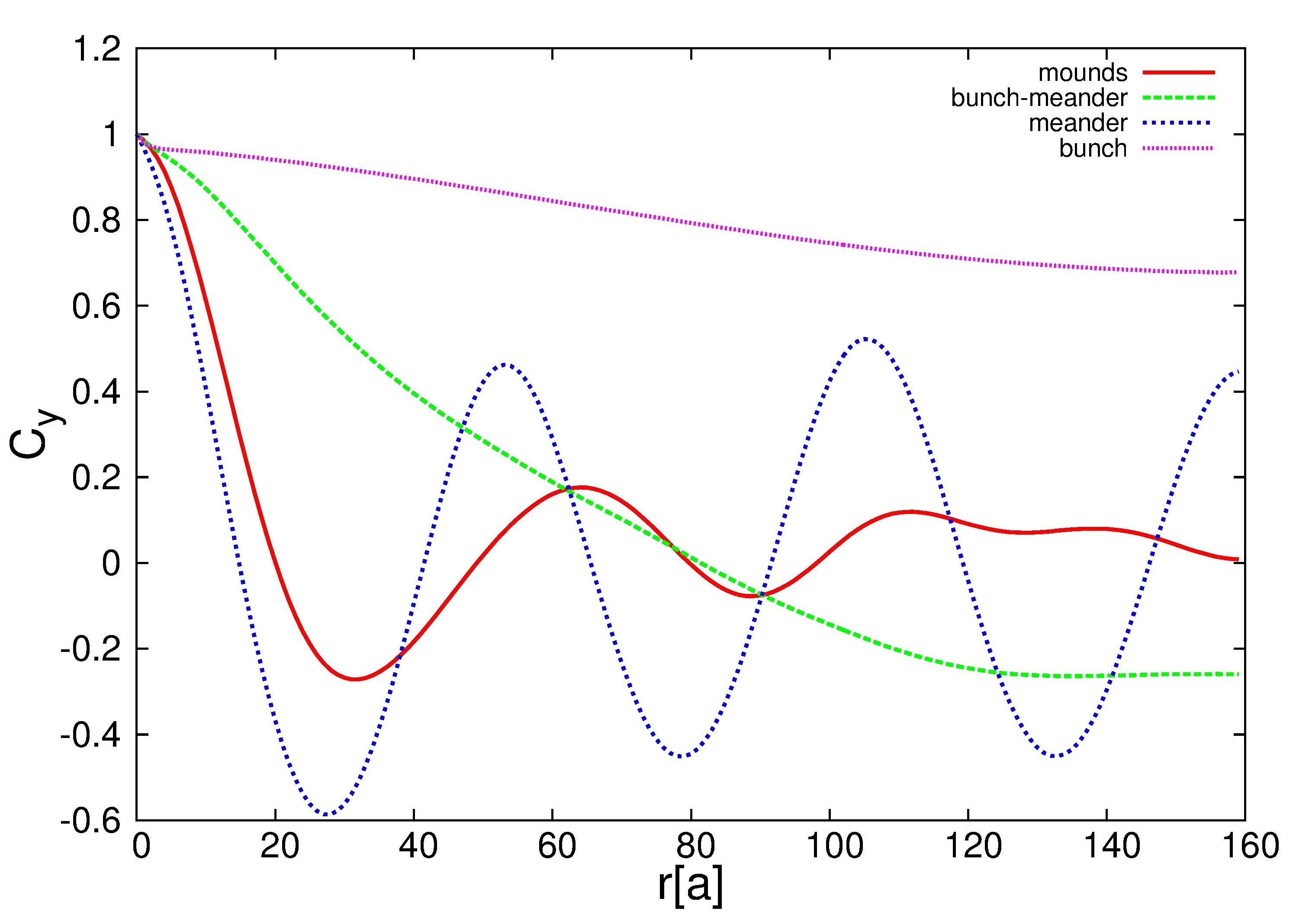}
\caption{Correlation function along $y$ direction for the same surfaces as in Fig. 7 plotted at $t=10^7$ \label{cy}}
\end{figure}
\begin{figure}
\includegraphics[width=8cm]{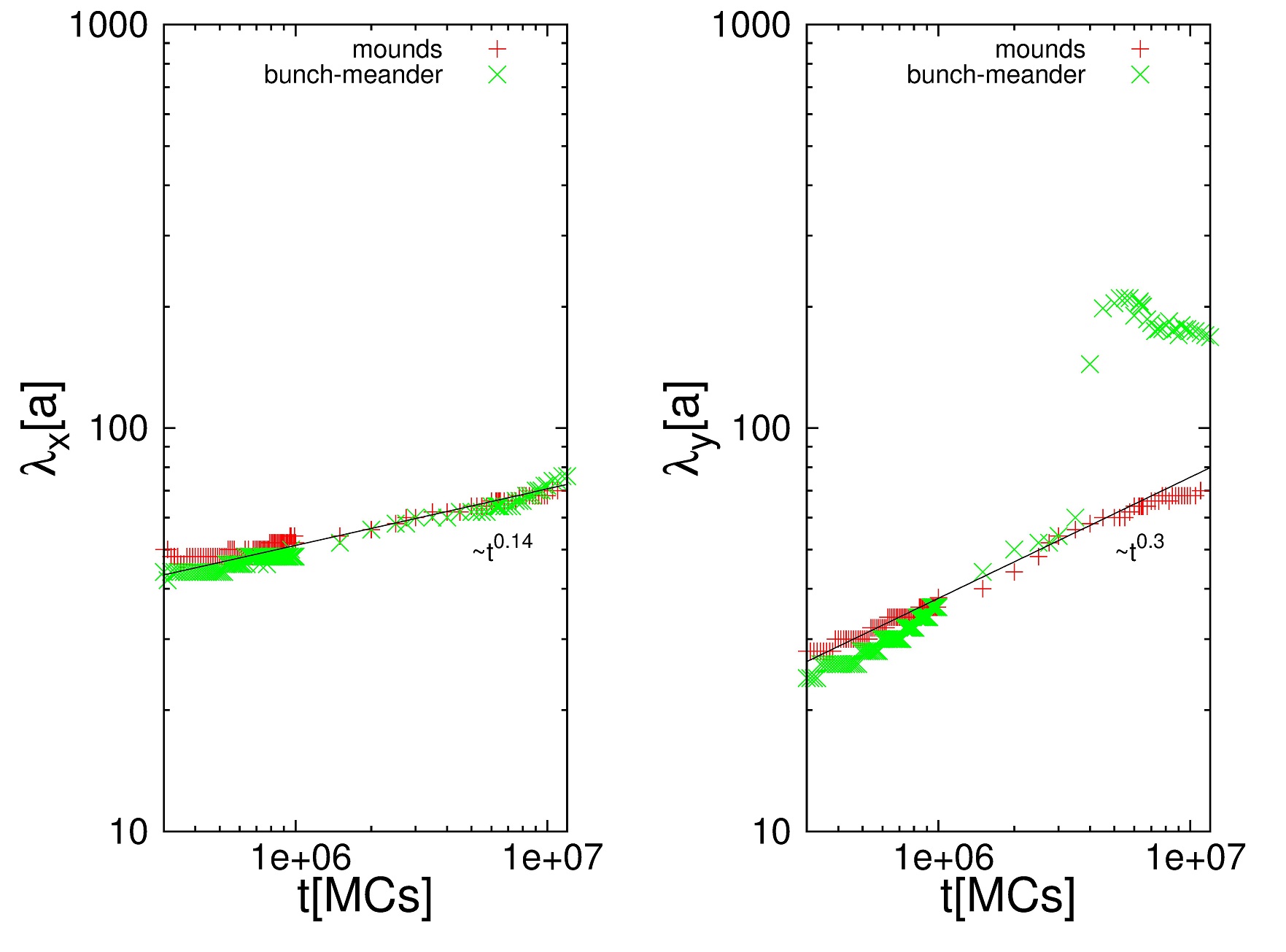}
\caption{Characteristic wavelength along axis x and y for mounded (Fig 5) and meandered and bunched (Fig 6) structures.
\label{length}}
\end{figure}
\begin{figure}
\includegraphics[width=10cm]{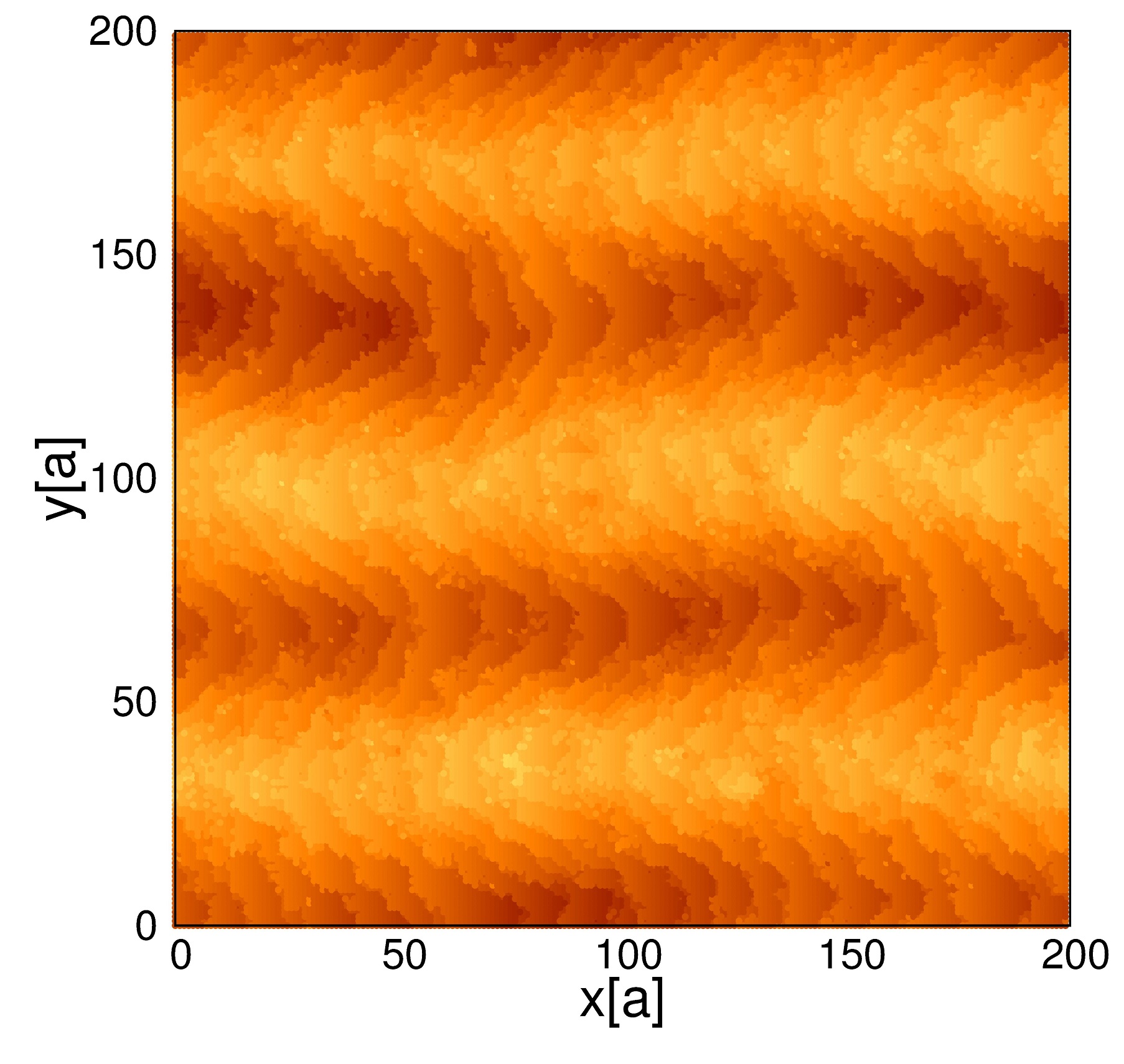}
\caption{\label{meander} Meandered  structure at SiC surface for $\ln(F/D)=-12.6$ and $\beta=6$.}
\end{figure}
Correlation functions for both structures can be also compared in Figs. \ref{cx} and \ref{cy}. Correlation function is calculated as the height correlation along $x$  axis $C_x=\sum_i (h_i-h_0) (h_{i-x}-h_0)$ and along  $y$ axis $C_x=\sum_i (h_i-h_0) (h_{i-x}-h_0)$. Correlation function for  bunch and meander structure oscillates in the direction $x$  - in the direction of  step movement and shows no special structure along step direction. Oscillations in 
x direction are similar to these seen for correlation function of surface with bunches.  When the same  correlation functions are calculated for mounds oscillations of the same length are present in both directions. They are  damped due to the  mound size and different location.  Time dependence of the correlation length measured as the position of first minimum of correlation function is shown in Fig \ref{length}.  It can be seen that up to the $ 4^.10^6 MC $ time steps   points for  mounds and bunch and meandered structures are at one curve. It means that step deformation along steps and across steps happens in the same way. Afterwards the wavelength along steps (y axis)  for this second structure jumps up thus showing emergence of elongated structures in this directions - bunches.

 Coexistence of bunching and meandering instability was shown in Ref.\onlinecite{neel} at metallic vicinal surfaces and several possible mechanisms were discussed there. Among these  mechanisms  ISB was excluded with argumentation that it leads to straight, not meandered steps due to its stabilizing  effect against meandering. This effect however is not so strong  when step movement together with the diffusion along steps  are not neglected in the stability analysis.
In Ref. \onlinecite{yu,yu2} simultaneous bunching and meandering was studying by use of phase-field model and 
deposition rate  was  a controlling parameter. This phenomenon was shown to emerge as collective SB and stress due to elastic interaction.  We assumed  in our simulations ISB and diffusion along surface and along steps  and interplay of these factors resulted in several different step patterns.

When  we  decrease particle flux  keeping  low temperatures meandered in phase  structure emerges \cite{misbah,bales,P-L,bena,zaluska2}. We can see this structure in Fig \ref{meander}. Regular meandered patterns are realized for realtively high ISB. The particle fluxes up the steps along step edges win over ISB surface  effect in this region of parameters. Roughness for this meandered pattern is much lower than the one seen in the   previous examples but correlation function in Fig. \ref{cy}  shows very large variability in the direction perpendicular to the step move direction. In the direction of  crystal miscut the same structure is very smooth (Fig \ref{cx}). Higher temperatures together with lower growth rates make step to straighten and eventually left down side part of the phase diagram is similar to this in the diagram in Fig \ref{ph2} occupied by four step structure, which again builds up for low $F$ value and high temperatures.

 Two different factors play role here: the temperature and  the growth rate. The system is two dimensional, so 1D models of step evolution   are not complete in this case. On lowering   temperature the importance of ISB barrier grows, what first of all prevents domain  creation at steps. Such effect can be  responsible for the transition from  mounds to  the meandered bunches for lower temperatures.
 It has to be  taken also  into account that  on changing temperature step permeability also changes what can have dramatic results \cite{stoyanov,sato1}. Moreover  the balance between kink and step edge energies changes what leads to the change of the  flux balance . An interplay of all of these factors results in different step patterns for different  model parameters.

\section{Conclusions}
Kinetic Monte Carlo simulations were done for Si(0001) surface with  $8^o$ miscut  along  $[01\bar10]$  direction.
Interplay between the particle flux imbalanced due to the step edge diffusion,   natural step movement and this caused by the presence of ISB was studied. These factors induce net particle fluxes moving in different directions. The value of flux difference induced by these factors changes with the step velocity  and temperature.  Step velocity depends on the crystal growth rate and the miscut angle. The importance of  ISB changes with the temperature of the system. The lower temperature is, the  ratio between different jumps over the surface is higher. On the other hand  for the higher  fluxes of   particles adsorbed at the surface relations between fluxes of all types can change. 

As  a result on changing growth condition different surface patterns are  observed,  similarly as in experimental situations. Temperature and growth rate are the usual parameters changed in experiments. Morphological instability can be also induced by  impurities i.e. by nitrogen \cite{ohtani,ohtani2}. We show that  if impurity influence onto the growth process can be treated  as the change in ISB height at steps it indeed explains emergence of surface structures.  
 Step bunching appears at rather low temperatures for  relatively high ISB. This  bunch structure is rather stiff and stable. When  temperature increases we get   the regular single step or four multi-step structure. For  lower  ISB  bunched and meandered structure appears  at low temperatures and at relatively high growth rate.  This pattern    changes into the  in phase meandered step  arrangement at lower external particle flux and into the regular mound structure at  higher temperatures. At  low growth rates and high temperatures steps at the surface  evolve towards straight four- step structure. Transition between different structures is sharp and the character  of  step pattern is easy to determine. 
In general  for higher temperatures and lower growth rates steps  become more straight.  

Similar phase diagrams are expected for other surface  miscuts. However because step velocity increases with lower miscut, transition lines  in these diagrams should be located  at different flux values.
                                                   
\section{Acknowledgement}
 This work was partially  supported by research grants from the National Science Centre(NCN) of Poland (Grant NCN No. 2011/01/B/ST3/00526) and from the  European Regional Development Fund, through grant Innovative Economy (POIG.01.01.02-00-008/08)

\end{document}